\documentclass[twocolumn,superscriptaddress,aps,prl,showpacs,floatfix]{revtex4}
\usepackage{hyperref}
\usepackage[sort&compress]{natbib}
\usepackage{amsmath,amssymb}
\usepackage{graphicx}

\def\figscalingfactorA{0.48}
\newcommand{\sech}{\operatorname{sech}}

\def\be{\begin{equation}}
\def\ee{\end{equation}}

\begin{document}

\date{\today}
\author{Nikolaos~K.~Efremidis}
\affiliation{Department of Applied Mathematics, University of Crete, 71409 Heraklion,
Crete, Greece}
\affiliation{School of Electrical and Computer Engineering, National Technical University
of Athens, Athens 15773, Greece}
\author{Kyriakos~Hizanidis}
\affiliation{School of Electrical and Computer Engineering, National Technical University
of Athens, Athens 15773, Greece}
\author{Boris~A.~Malomed}
\affiliation{Department of Interdisciplinary Studies, Faculty of Engineering, Tel Aviv
University, Tel Aviv 69978, Israel}
\author{Paolo~Di~Trapani}
\affiliation{Istituto Nazionale per la Fisica della Materia and Department of Physical
and Mathematical Sciences, University of Insubria, Via Valleggio 11, 22100
Como, Italy}
\title{Three-dimensional vortex solitons in self-defocusing media}

\begin{abstract}
We demonstrate that families of vortex solitons are possible in a
bi-dispersive three-dimensional nonlinear Schr\"odinger equation. These
solutions can be considered as extensions of two-dimensional dark vortex
solitons which, along the third dimension, remain localized due to the
interplay between dispersion and nonlinearity. Such vortex solitons can
be observed in optical media with normal dispersion, normal diffraction, and
defocusing nonlinearity.
\end{abstract}

\maketitle

Vortex solitons are self-localized solutions of nonlinear wave equations,
which are characterized by a phase singularity at the pivotal point. The
phase charge of a simple closed curve surrounding the vortex core is equal
to $2\pi m$, where $m$ is the integer vorticity of the solution.
Vortex solitons have been theoretically predicted in the context of
superfluids~\cite{ginzb-jetp1958,pitae-jetp1961}. In those early works, a
two-dimensional (2D) nonlinear Schr\"{o}dinger (NLS) equation with
defocusing nonlinearity was shown to support vortex soliton solutions whose
intensity vanishes at the vortex center and asymptotically approaches a
constant value at infinity. Such dark vortex solitons were experimentally
observed in a bulk self-defocusing optical medium~\cite{swart-prl1992}. The
stability of nonlinear vortices depends on the vorticity number, $m$.
Fundamental vortices with $m=1$ are energetically favorable (and, as a
result stable), which implies that instabilities of other families of
solutions may result to the formation of (a set of) fundamental vortices. In
particular, a 2D dark soliton stripe is unstable to long-wave symmetry
breaking perturbations, leading to the generation of fundamental vortex
soliton pairs with opposite vorticities~%
\citep{law-ol1993,mamae-prl1996,tikho-ol1996}. Higher order vortices are
also unstable and break down into fundamental ones. However, in the NLS
limit no exponentially growing mode exists~\cite{arans-prb1996} (the
instability may be \textit{subexponential}) and, as a result, multicharged
vortices are very long-lived objects. Strong instabilities of
multicharged vortices can be triggered by different mechanisms such as
dissipation, nonlinearity saturation, or anisotropy~\cite{mamae-prl1997}.

Another class of solutions feature ring-shaped intensity profiles
and exist in self-focusing media~\cite{krugl-pla1985}. Such solutions are
unstable even in the case of saturable nonlinearity due to
azimuthal instabilities and break down to a set of fundamental
solitons~\cite{tikho-josab1995}. Ring vortices can be stabilized if the
model includes a combination of competing (self-focusing and
self-defocusing) nonlinear terms, such as cubic and quintic \cite{CQ,Pramana}
or quadratic and self-defocusing cubic \cite{chi2chi3} ones. Azimuthal
instabilities can also be suppressed by appropriately modulating
the amplitude of the solution in the angular direction~\cite{desya-prl2005}.
Recently a lot of attention has been attracted to the study of vortices in
periodic lattices. The lattice can stabilize families of ring
vortices~\cite%
{malom-pre2001,BBB,yang-ol2003,fleis-prl2004,neshe-prl2004-vortex} by
trapping the intensity in the lattice potential minima. In a
different setting, vortex solitons were observed in Bose-Einstein
condensates~\cite{ginsb-prl2005,ander-prl2001}.

In contrast to the plethora of theoretical and experimental works on
vortices in two dimensions, only a few works have addressed 3D solutions.
Stable toroidal solitons with vorticity $1$ were found in systems with
competing nonlinearities~\cite{mihal-prl2002,chi2chi3}, and a variety of 3D
discrete solitons of the vortex type were constructed and explored
in Refs. \cite{3D-DNLS1,3D-DNLS2}. In addition, propagation and robustness
of 3D ring optical vortices in the atmosphere was examined in~Ref.~\cite{vinco-prl2005}.
Focusing properties of bi-dispersive (normal
dispersion and normal diffraction) optical systems have been studied in~\cite{berge-josab1996,berge-pre1996} for self-focusing nonlinearities.

The subject of the present paper is to find 3D counterparts of the
dark optical vortices which were discovered long ago in~Refs. \cite%
{ginzb-jetp1958,pitae-jetp1961}, and to predict experimental conditions
necessary for their observation. We demonstrate that 3D vortices exist in
media with the cubic defocusing nonlinearity and normal GVD (group velocity
dispersion), i.e., the diffraction-dispersion operator is of the hyperbolic
type. Optical media realizing this model are available, such as specific
AlGaAs alloys~\cite{dumai-ol1996,gorza-prl2004}. In the transverse $\left(
x,y\right) $ plane, these solutions have the form of a dark vortex, whereas
along the longitudinal axis (i.e., in the temporal direction) they remain
localized as bright temporal solitons, due to the interplay between the
normal GVD and defocusing nonlinearity. We first construct the
solutions in a semi-analytical (and quite accurate) form, making use of the
Hartree approximation~\cite{hayat-pre1993}. Then, we employ the Newton
iteration method to find the 3D vortices as numerical solutions to the
underlying NLS equation. The Hartree approximation provides the Newton's
method with appropriate initial conditions. The stability of the
vortices is tested by direct numerical simulations. We conclude that 3D
vortices are stable (as long as a super-Gaussian carrying the vortex in the
transverse plane, which is as a part of the numerical procedure, does not
suffer essential diffraction). Considering applications, the
central hole in such a vortex pancake may be used as an optically induced
aperture to dynamically control a probe beam passing through it.

We start the analysis by introducing the normalized NLS equation,
\begin{equation}
i\psi _{z}+\frac{1}{2}\left( \nabla _{\perp }^{2}\psi -\psi _{tt}\right)
-|\psi |^{2}\psi =0,  \label{model}
\end{equation}%
where $\nabla _{\perp }^{2}=\partial _{x}^{2}+\partial _{y}^{2}$ is the
diffraction operator, with the normal GVD dispersion and defocusing
nonlinearity coefficients scaled to unity. In addition to the
aforementioned optical media (AlGaAs alloys), Eq.~(\ref{model}) applies, as
the Gross-Pitaevskii equation, to Bose-Einstein condensate in an optical
lattice, where $z$ is now time, and $t$  \ a
spatial coordinate across the lattice. In this case, the negative
diffraction along $t$ is achieved when the Bloch momentum corresponds to a
negative effective mass.

Notice that if $\psi (x,y,z,t)$ is a solution of Eq.~(\ref{model}), then $%
\psi ^{\prime }=\alpha \psi (\alpha x,\alpha y,\alpha t,\alpha ^{2}z)$,%
with real free parameter $\alpha$, is a solution too.
Using this scale-invariance property, a one-parameter family of
solutions can be generated from a single solution of Eq.~(\ref%
{model}). Introducing polar coordinates $(\rho ,\phi )$ in the $(x,y)$
plane, we look for solutions of Eq.~(\ref{model}) in the form of
\begin{equation}
\psi (\rho ,\phi ,t,z)=u(\rho ,t)\exp (-ikz)\exp (im\phi ).  \label{ansatz}
\end{equation}%
where $m$ is integer vorticity. Substituting Eq.~(\ref%
{ansatz}) in Eq.~(\ref{model}) and using the above-mentioned scale
invariance to set $k=1/2$, we obtain:
\begin{equation}
u+\left( \rho ^{-1}u_{\rho }+u_{\rho \rho }-m^{2}\rho ^{-2}u\right)
-u_{tt}-2u^{3}=0.  \label{model_static}
\end{equation}

The asymptotic expansion of the $t$-independent problem at $\rho
\rightarrow \infty $ yields $u=(1/\sqrt{2})(1-(m^{2}/2)\rho
^{-2}-(m^{2}/8)(8+m^{2})\rho ^{-4})+\mathcal{O}(\rho ^{-6})$. On the other
hand, the asymptotic expansion at $\rho \rightarrow 0$ is
$u=c(\rho ^{|m|}-(1/(4(m+1))\rho ^{|m|+2})+\mathcal{O}(\rho ^{|m|+4})$.
Accordingly, we look for solutions of the time-dependent problem
with $\lim_{\rho \rightarrow
0}u(\rho ,t)=0$ and $\lim_{\rho \rightarrow \infty }u(\rho
,t)=u_{\infty }(t)$. In the latter limit, Eq.~(\ref{model_static})
reduces to the dynamical system associated with the single-soliton 
solution of the NLS equation, $u_{\infty }-u_{\infty ,tt}-2u_{\infty }^{3}=0$, whose commonly known soliton solution is
\begin{equation}
\lim_{\rho \rightarrow \infty }u(\rho ,t)=u_{\infty }(t)=\sech(t),
\label{u_infty}
\end{equation}%
in compliance with our objective to find solutions that look as bright
solitons in the temporal direction.

The total energy conserved by Eq.~(\ref{model}) is
\begin{equation}
P=\int_{-\infty }^{\infty }\int_{-\infty }^{\infty }\int_{-\infty }^{\infty
}|\psi (x,y,t)|^{2}\,dx\,dy\,dt.  \label{p}
\end{equation}%
However, this integral of motion diverges for solutions with
asymptotic form (\ref{u_infty}). A renormalized (convergent) form
of the energy, that takes into account the asymptotic form of the solution,
may be defined as
\begin{equation}
P_{n}=\int_{-\infty }^{\infty }\int_{-\infty }^{\infty }\int_{-\infty
}^{\infty }(u_{\infty }^{2}(t)-|\psi |^{2})\,dx\,dy\,dt.  \label{p_n}
\end{equation}%
Notice that $\psi _{s}=u_{\infty }(t)e^{iz/2}=\sech(t)e^{iz/2} $ is a solution of Eq.~(\ref{model}) representing a
bright soliton stripe in three dimensions. As a result, both terms in~Eq. (%
\ref{p_n}) obey the above-mentioned scale invariance, which can be
used to derive a rule for generating a one-parameter family of
vortex solitons from a single one: $P_{n}(\alpha )=P_{n}(1)/\alpha $%
.

We start the analysis by resorting to the Hartree approximation
(HA), which is based on the product ansatz~\cite{hayat-pre1993},
\begin{equation}
u(\rho ,t)=R(\rho )T(t).  \label{hartree}
\end{equation}%
This approximation will then be used as an initial guess for
numerical solutions based on the Newton's method. Substituting Eq.~(%
\ref{hartree}) in~Eq. (\ref{model_static}), we arrive at a formal
equation,
\begin{equation}
\left( \rho ^{-1}R^{\prime }+R^{\prime \prime }-m^{2}\rho ^{-2}R+R\right)
T-RT^{\prime \prime }-R^{3}T^{3}=0  \label{tmp}
\end{equation}

First, Eq.~(\ref{tmp}) is to be multiplied with $R$, integrated
over $\rho $ from $0$ to $\rho _{1}$, and divided by $\int_{0}^{\rho
_{1}}R^{2}\,d\rho $. Taking the limit of $\rho _{1}\rightarrow \infty $, the
equation
\begin{equation}
T-T^{\prime \prime }-2T^{3}=0.  \label{T}
\end{equation}%
is derived by assuming (without loss of generality) that $R(\rho \rightarrow
\infty )=1$. A relevant solution to Eq.~(\ref{T}) is the same
temporal bright soliton as the one obtained above as the asymptotic
wave form, $T(t)=\sech(t)$. Next, we multiply Eq.~(\ref{tmp}) by $T$,
integrate it from $-t_{1}$ to $t_{2}$, divide by $%
\int_{-t_{1}}^{t_{2}}T^{2}(t)\,dt$, and take the limits of $t_{1}\rightarrow
\infty ,t_{2}\rightarrow \infty $, to obtain
\begin{equation}
3\left( \rho ^{-1}R^{\prime }+R^{\prime \prime }-m^{2}\rho ^{-2}R\right)
+4(R-R^{3})=0.  \label{radial}
\end{equation}%
Equation~(\ref{radial}) can be solved numerically by dint of
standard two-point boundary-value methods, such as shooting.
\begin{figure}[tbp]
\centerline{\includegraphics[width=\figscalingfactorA\columnwidth]{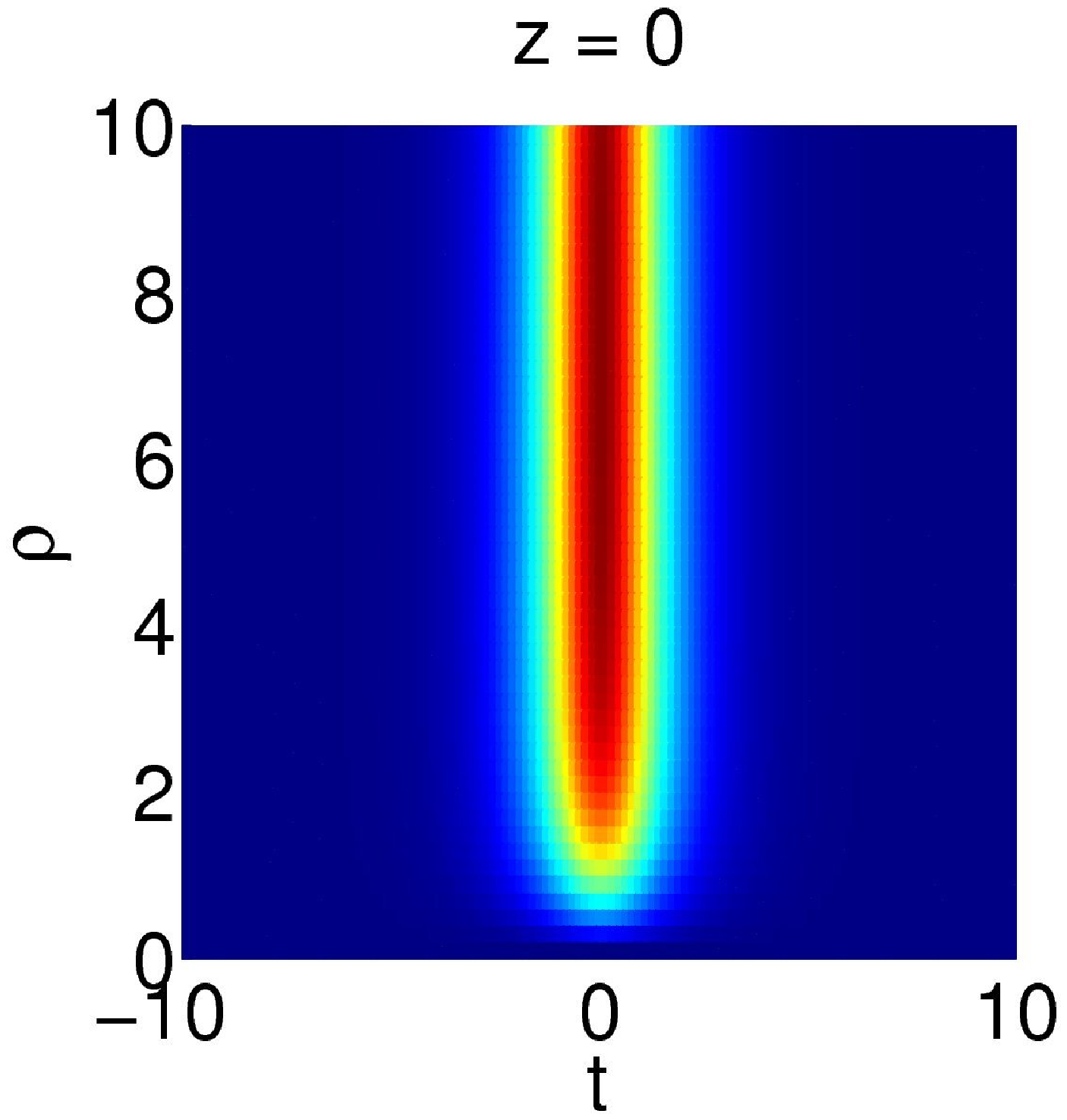}%
\includegraphics[width=\figscalingfactorA\columnwidth]{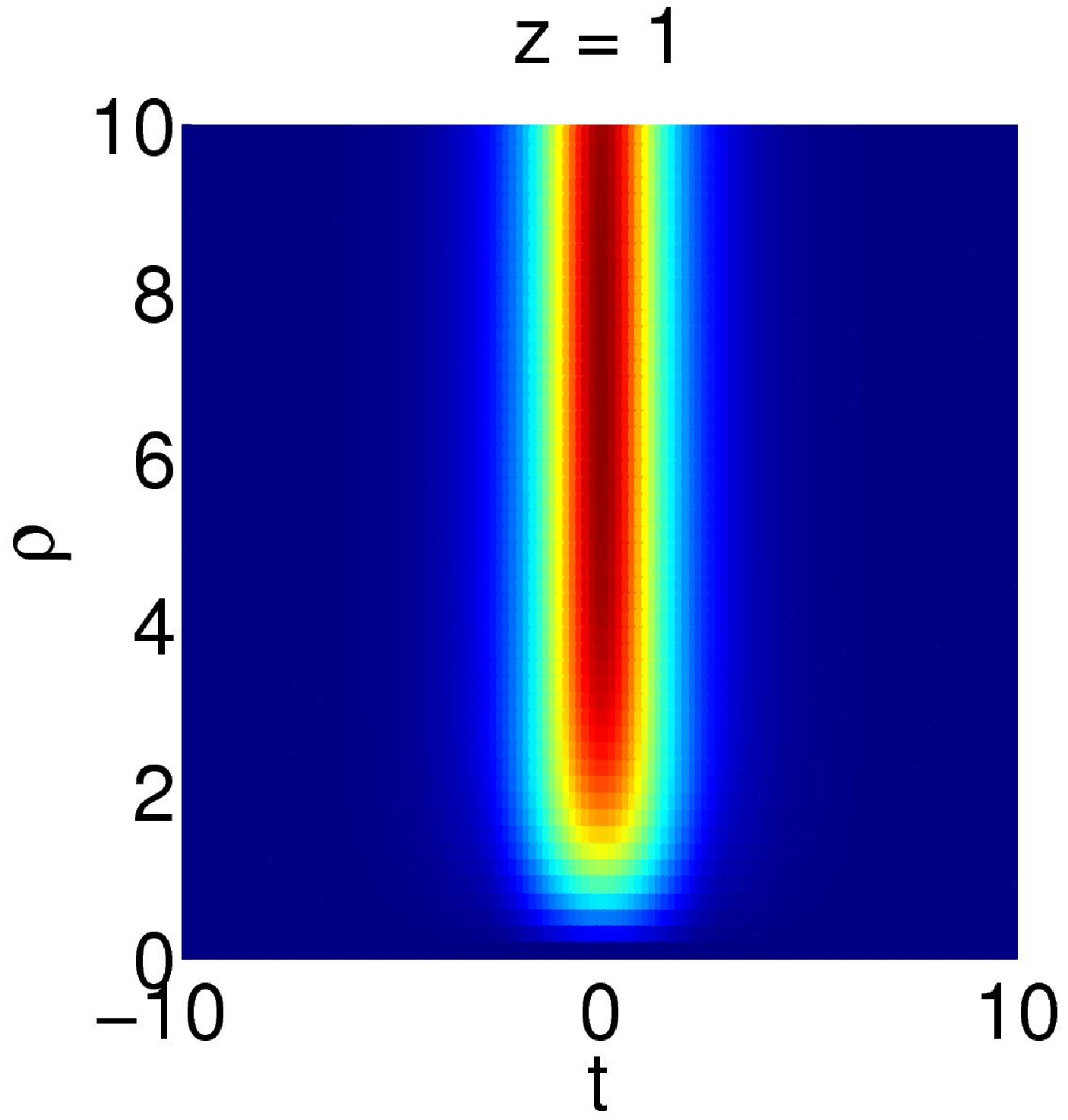}} %
\centerline{\includegraphics[width=\figscalingfactorA\columnwidth]{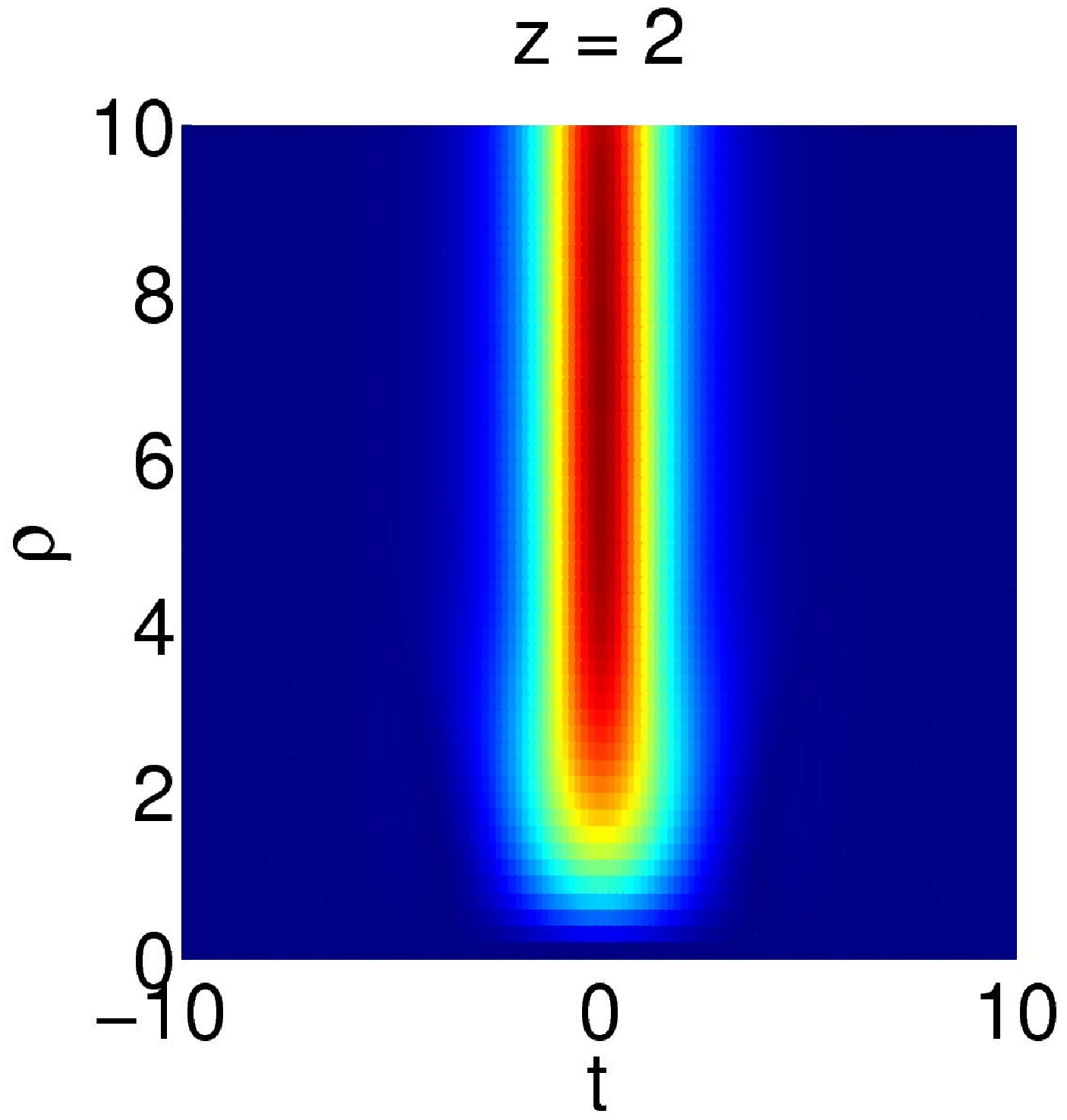}%
\includegraphics[width=\figscalingfactorA\columnwidth]{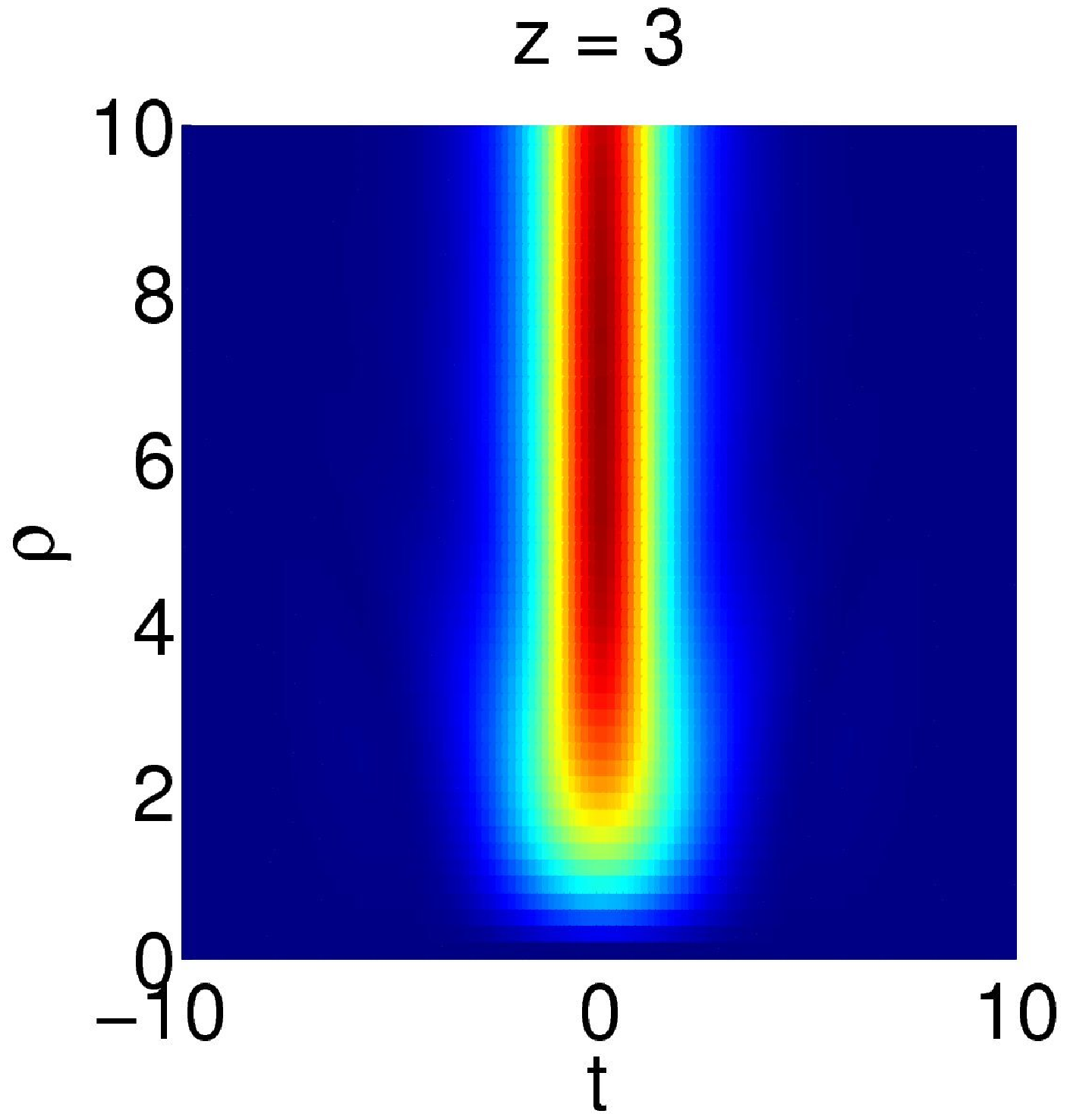}}
\caption{Shape of the fundamental ($m=1$) vortex soliton solution, at $%
z=0,1,2,3$. The initial condition is taken as per the Hartree
approximation. Here and below, contour plots display the wave
amplitude, $|\protect\psi |$, versus $t$ and $\protect\rho $.}
\label{fg:hartree_soliton_bpm}
\end{figure}

All our direct numerical simulations are performed using the beam propagation
method, where the linear part is solved using the fast Fourier transform and
the nonlinear part by direct integration (see, for example~\cite{newel-1992}). Since the solution does not vanish at infinity, and aiming to avoid
artifacts produced by reflections from domain boundaries, we took a 
$12$-th order super-Gaussian (in $\rho $) as a
finite-extension carrier for the solution. Of course, diffraction of the
super-Gaussian cannot be avoided, hence, after a large but finite
propagation distance, the field starts to decay. Eventually, this
leads to destabilization of the vortex.

In Fig.~\ref{fg:hartree_soliton_bpm} the evolution of an initial 
configuration suggested by the HA is depicted. Notice that the simulation
window is truncated for large values of $\rho $ so as to display only the
evolution of the vortex soliton while eliminating the above-mentioned
irrelevant effect of the background diffraction. In all
simulations, the initial conditions \emph{did not} develop azimuthal
instabilities, i.e., the intensity distribution corresponding to the
vortex soliton did not generate any dependence on $\phi $, nor did
it  develop any other instability. Furthermore, as the vortex propagates,
it slightly broadens in $t$ close to its center. This effect does
not imply any trend to decay of the vortex. It is, rather, a
manifestation of the relaxation of the initial approximate wave form towards
an exact vortex state. As shown below, this conclusion is in
agreement with results generated by the Newton iteration method.

\begin{figure}[tbp]
\centerline{\includegraphics[width=0.96%
\columnwidth]{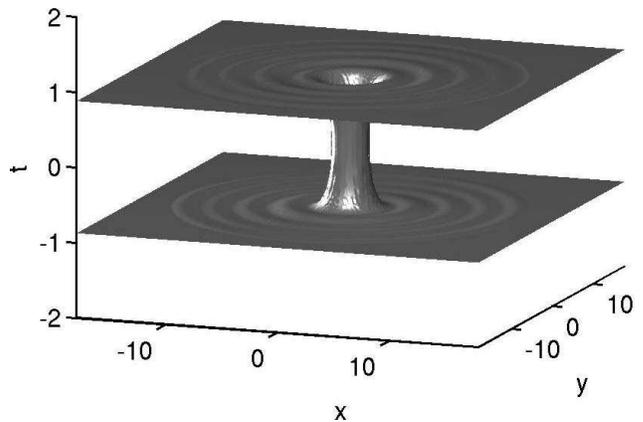}}
\caption{Isosurface plot, corresponding to $\left\vert u\right\vert ^{2}=0.5$%
, of the fundamental vortex soliton solution obtained by means of the Newton
iteration method.}
\label{fg:newton_ic}
\end{figure}

To generate numerically exact stationary vortex-soliton solutions,
the HA was fed, as an initial guess, into the Newton's method for Eq.~(\ref%
{model_static}). Notice that the discretization of Eq.~(\ref{model_static})
in variable $\rho $ requires special attention, since the error is $\rho$%
-dependent. More specifically, the discretization error becomes
larger as $\rho $ decreases, whereas at large values of $\rho $
the error is almost isotropic (independent of $\rho $, to the leading
order). The iteration provided for the convergence of solutions
fed by the above-mentioned initial HA configurations. The
iso-intensity profile of the so obtained fundamental vortex soliton is
depicted in Fig.~\ref{fg:newton_ic}.

\begin{figure}[b]
\centerline{\includegraphics[width=\figscalingfactorA\columnwidth]{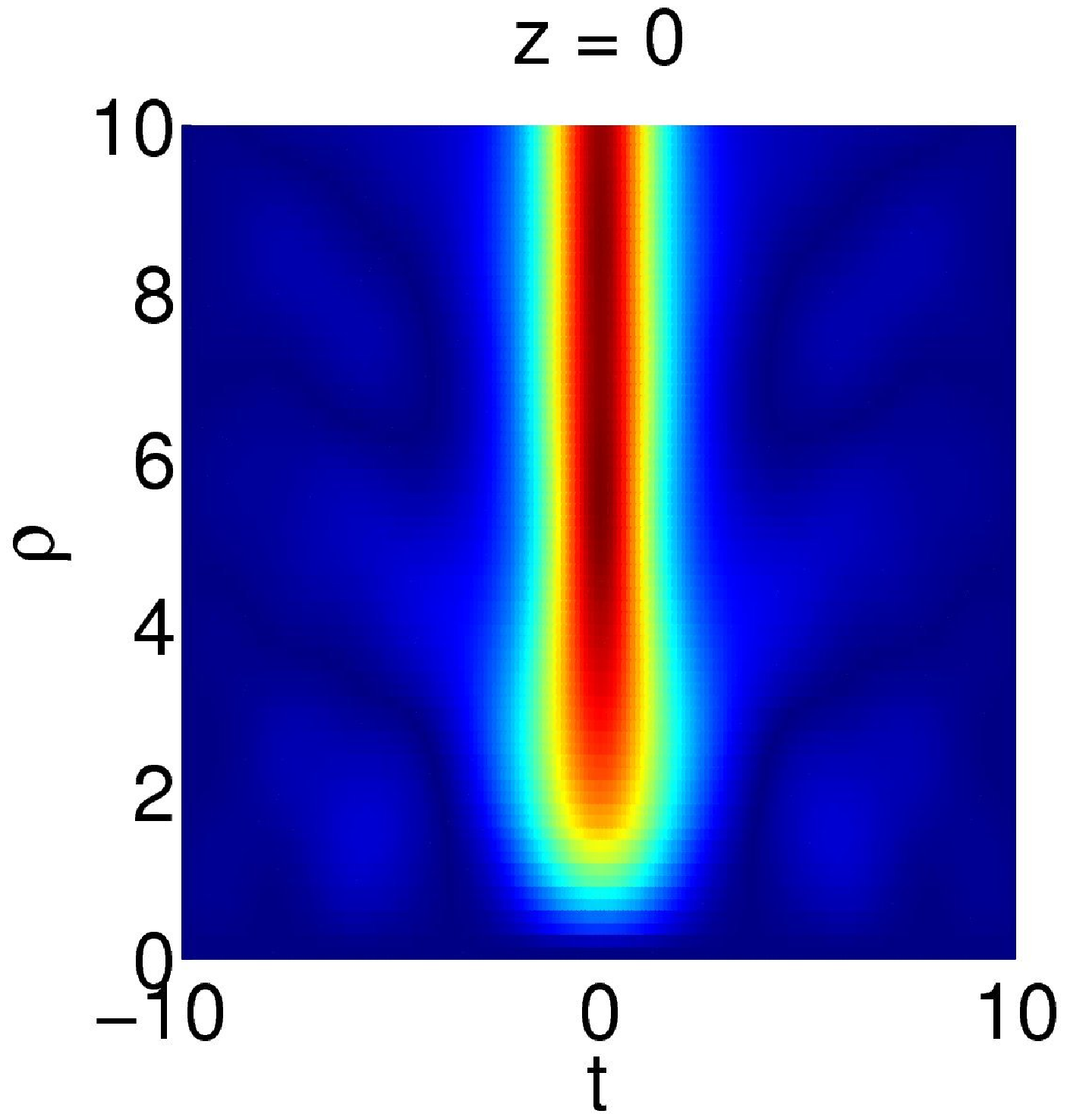}%
\includegraphics[width=\figscalingfactorA\columnwidth]{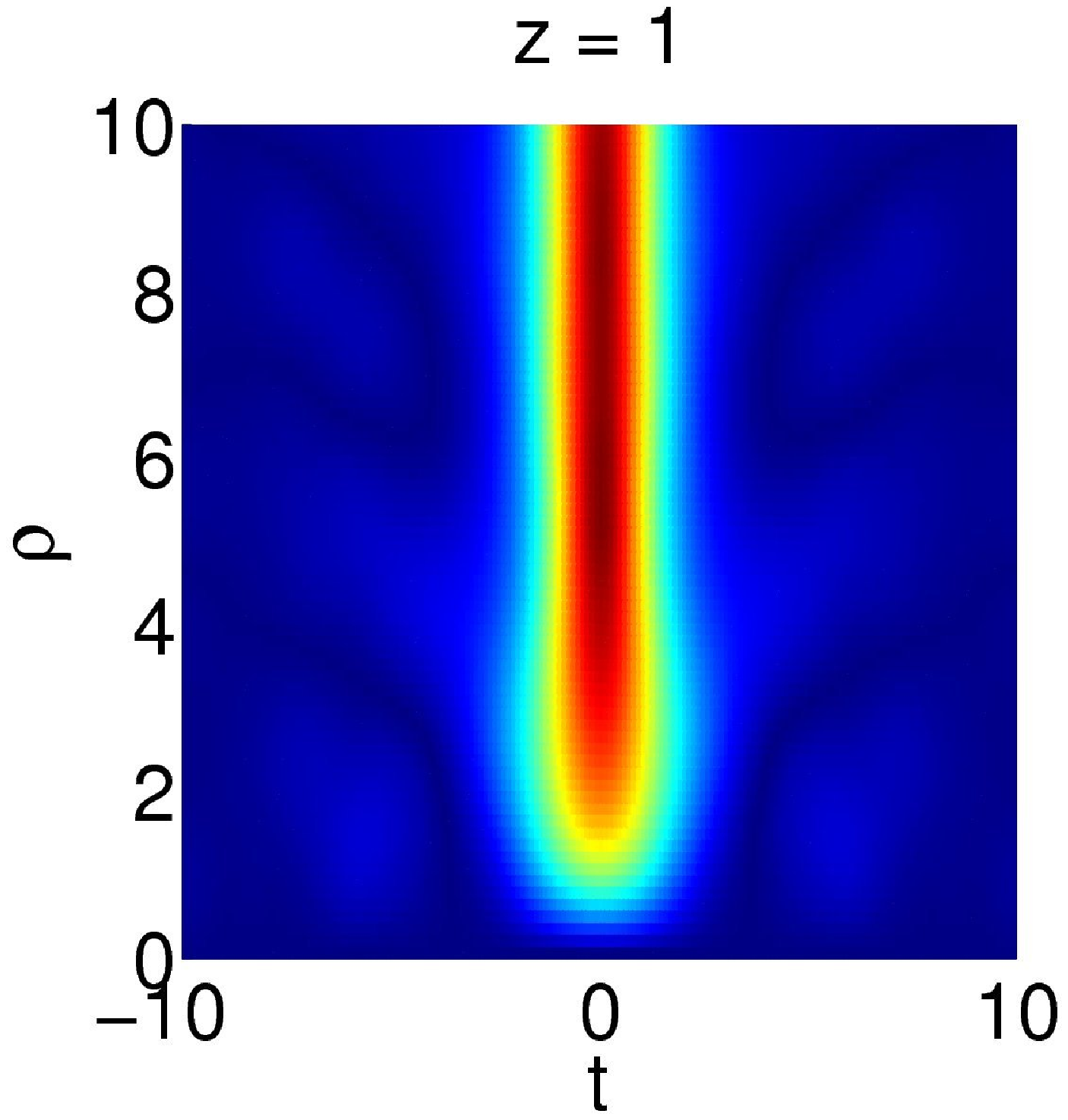}} %
\centerline{\includegraphics[width=\figscalingfactorA\columnwidth]{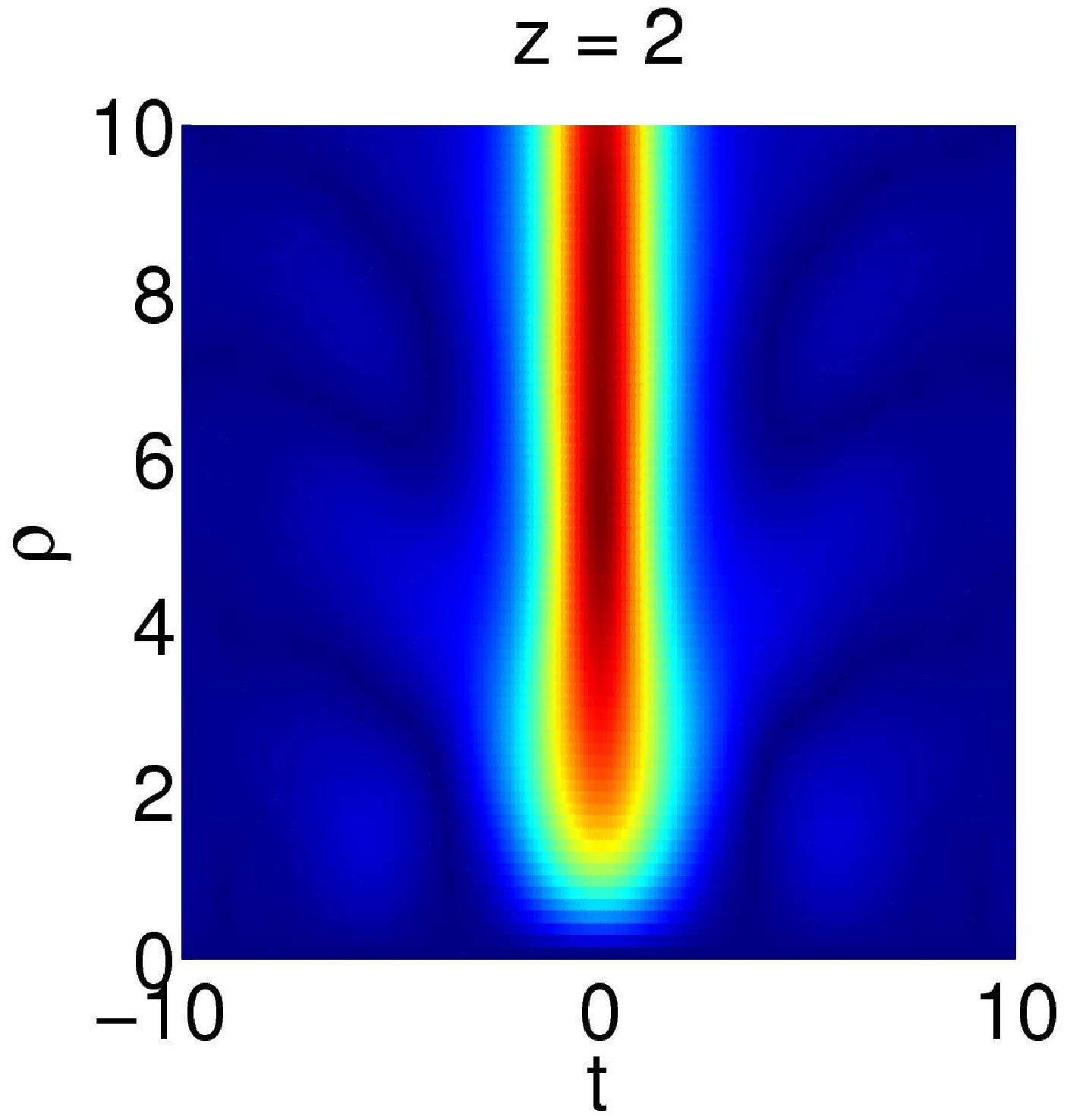}%
\includegraphics[width=\figscalingfactorA\columnwidth]{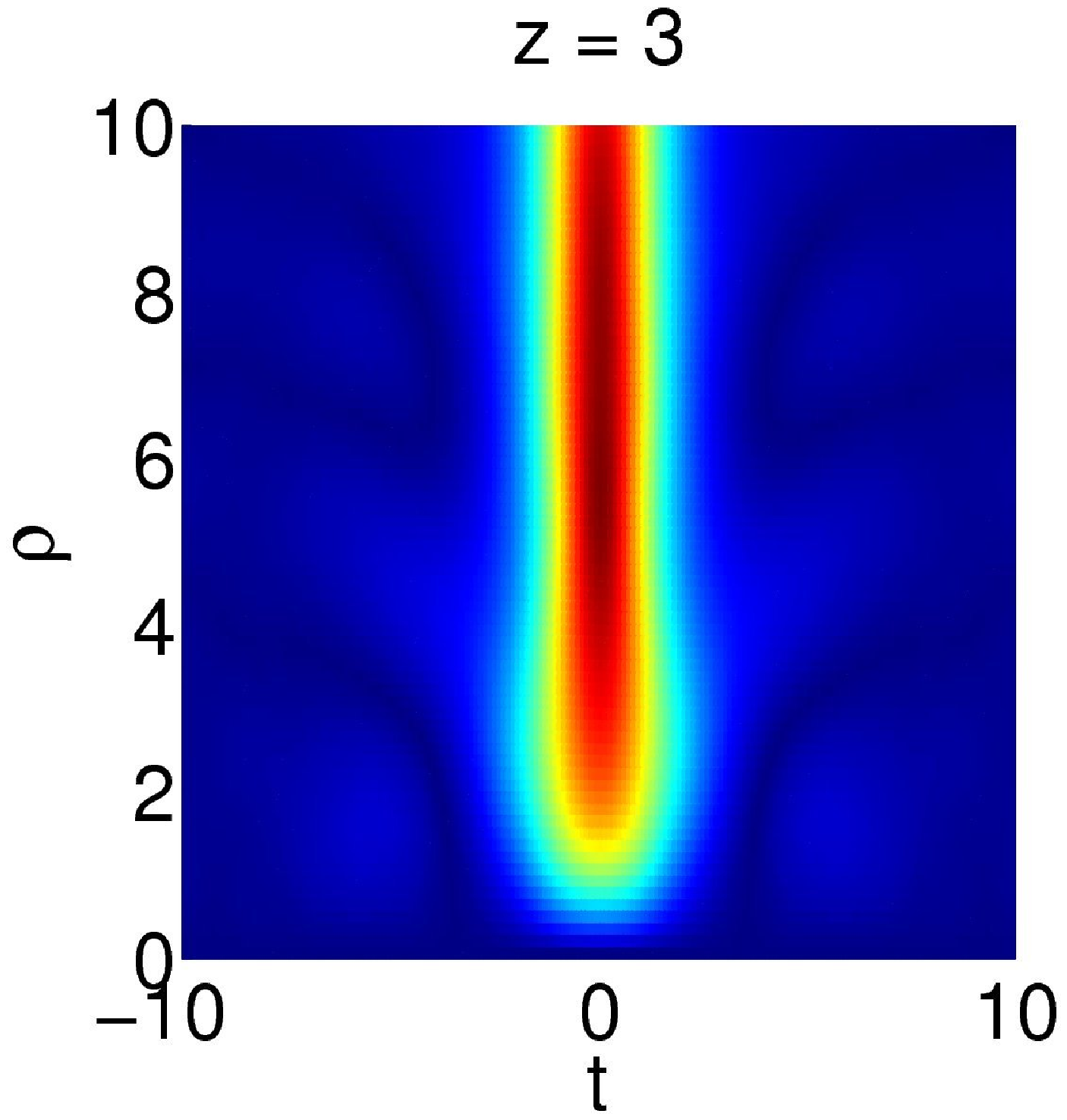}}
\caption{The same as in Fig.~\protect\ref{fg:hartree_soliton_bpm}, but if
the initial condition was taken as a stationary solution generated
by the Newton's iteration method.}
\label{fg:newton_soliton_bpm}
\end{figure}

\begin{figure}[tbp]
\centerline{\includegraphics[width=\figscalingfactorA\columnwidth]{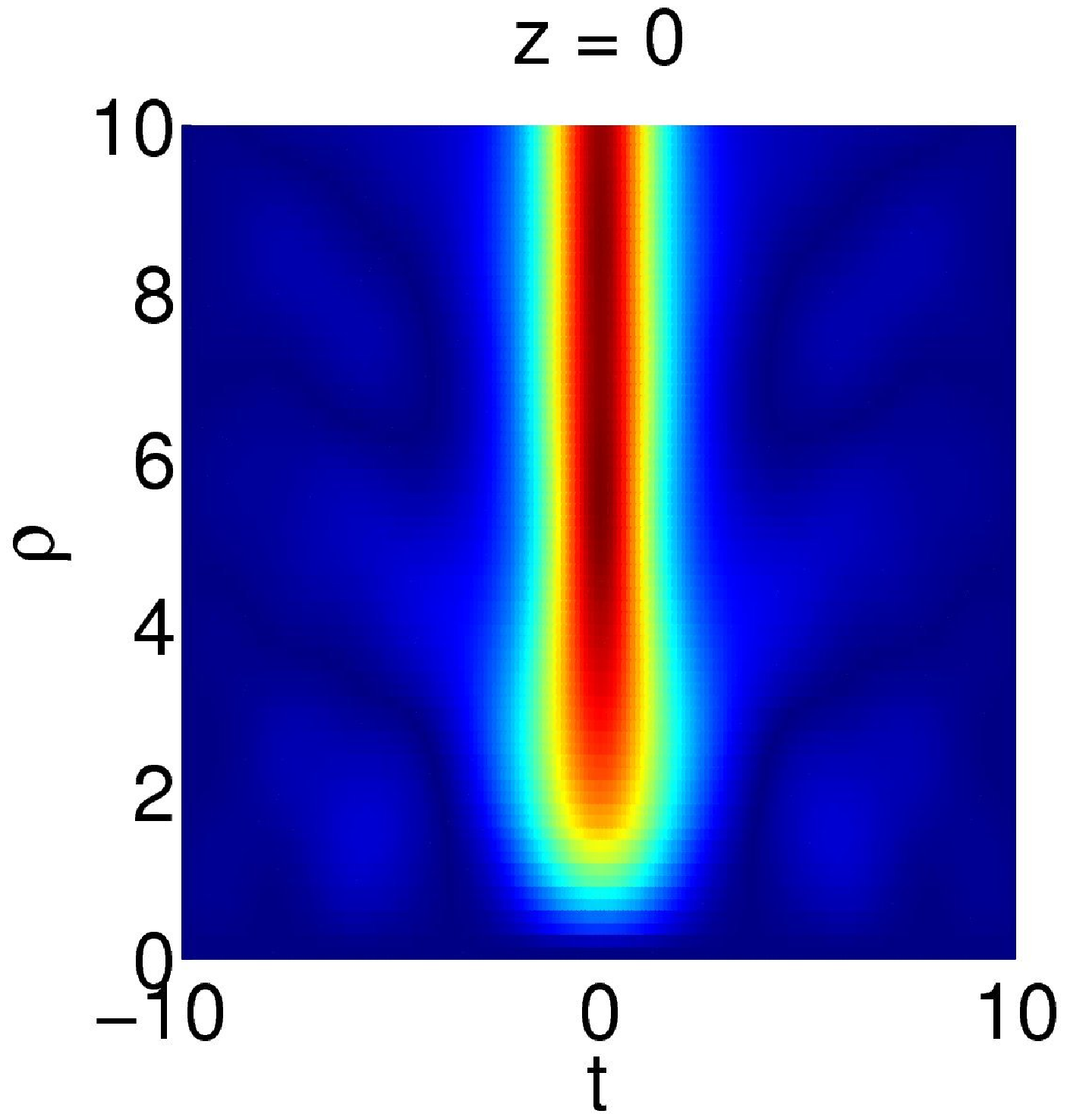}%
\includegraphics[width=\figscalingfactorA\columnwidth]{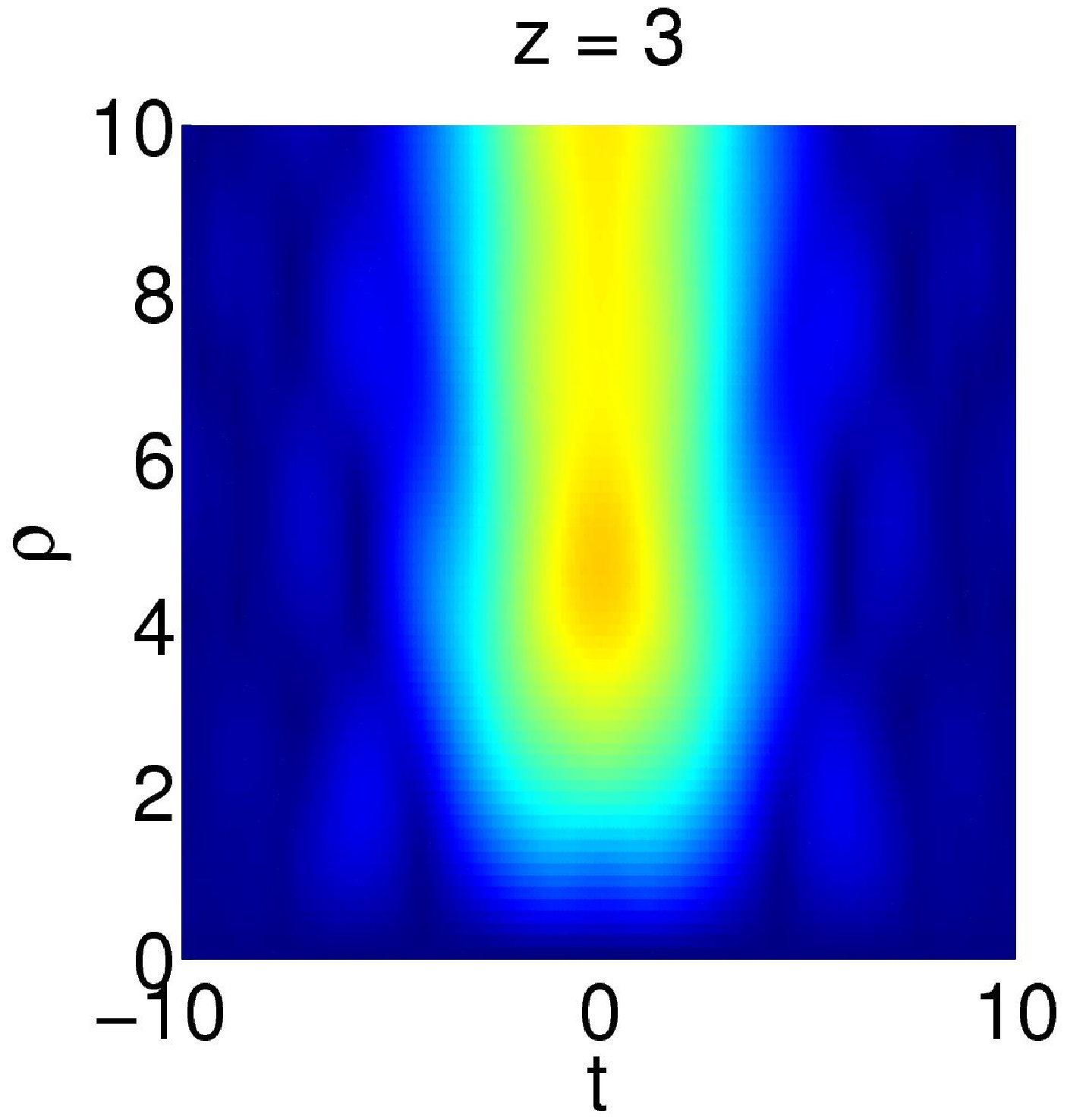}}
\caption{Propagation of a fundamental ($m=1$) vortex soliton up to $z=0,3$
at which point the nonlinearity is switched off.}
\label{fg:newton_diffraction_bpm}
\end{figure}

In Fig.~\ref{fg:newton_soliton_bpm}, direct simulations of the
propagation of a 3D vortex soliton, with the initial condition
generated by the Newton's iteration method, is shown (as above, the initial
configuration was actually multiplied by the $12$-th order super-Gaussian).
Notice that the vortex soliton profile remains invariant in the
course of the propagation, i.e., the vortex soliton is \emph{stable}.
Comparison with Fig.~\ref{fg:hartree_soliton_bpm}, which showed similar
evolution initiated by the HA, attests to the accuracy of that
approximation. A noteworthy feature revealed by Fig.~\ref%
{fg:newton_soliton_bpm} (in a more salient form than by the HA, cf. Fig.~\ref%
{fg:hartree_soliton_bpm}) is that, close to the vortex core (at small values
of $\rho $), the vortex is wider in the $t$-direction. We stress
that the vortical phase structure of the solution is maintained during the
propagation.

To directly verify that such a 3D vortex soliton solution is a
nonlinear object indeed, we repeated the same simulations, dropping the
nonlinear term. As one can see in Fig.~\ref{fg:newton_diffraction_bpm},
the vortex quickly diffracts in that case by at $z=3$.

We have also performed simulations for double vortices, with $m=2$.
They were found to propagate undistorted over large distances,\emph{%
\ without splitting} into fundamental vortices. This observation
is, in fact, in agreement with a conjecture put forward in Ref.~\cite{arans-prb1996}.

In conclusion, we have demonstrated that vortex solitons are possible in the
bi-dispersive 3D NLS equation. These solutions have the form of a dark
vortex in the spatial plane, whereas they are localized
along the temporal dimension, due to the interplay between the GVD and
nonlinearity. Such vortex solitons can be observed in certain optical media
with normal GVD, normal diffraction, and defocusing nonlinearity, and in
self-repulsive BEC.

N.K.E. and K.H. Acknowledge funding by the European Social Fund (75\%) and
National Resources (25\%)-Operational Program for Educational and Vocational
Training II (EPEAEK II) and particularly the Program PYTHAGORAS. B.A.M.
appreciates hospitality of the National Technical University of Athens.


\end{document}